\documentclass[fleqn,usenatbib]{mnras}

\usepackage{newtxtext,newtxmath}
\usepackage[T1]{fontenc}
\usepackage{ae,aecompl}

\usepackage{graphicx}	% Including figure files
\usepackage{amsmath}	% Advanced maths commands
\usepackage{amssymb}	% Extra maths symbols

\title[Lensing efficiency for gravitational wave mergers]{Lensing efficiency for gravitational wave mergers}

\author[O. Contigiani]{
O. Contigiani\thanks{E-mail: contigiani@strw.leidenuniv.nl}
\\
Leiden Observatory, Leiden University, PO Box 9506, Leiden 2300 RA, The Netherlands \\
Lorentz Institute for Theoretical Physics, Leiden University, PO Box 9506, Leiden 2300 RA, The Netherlands
}

\date{Accepted XXX. Received YYY; in original form ZZZ}

\pubyear{2015}

\begin{document}
\label{firstpage}
\pagerange{\pageref{firstpage}--\pageref{lastpage}}
\maketitle

\begin{abstract}
We gain insight into the effects of gravitational lensing on the estimated distribution of merging binaries observed through gravitational waves. We quantify the efficiency of magnification for gravitational wave events in the geometric optics limit, and we compare it to the electromagnetic case by making minimal assumptions about the distribution of intrinsic properties for the source population. We show that lensing effects leave a recognizable signature on the observed rates, and that they can be prominent only in the presence of an extremely steep mass function (or redshift evolution) and mainly at low inferred redshifts. We conclude that gravitational magnification does not represent a significant systematic for gravitational wave merger studies in the LIGO-Virgo era.
\end{abstract}


%%%%%%%%%%%%%%%%%%%%%%%%%%%%%%%%%%%%%%%%%%%%%%%%%%

%%%%%%%%%%%%%%%%% BODY OF PAPER %%%%%%%%%%%%%%%%%%
\section{Introduction}
Even before the first detection of gravitational waves (GWs) due to the merger of a compact binary by the LIGO-Virgo collaboration \citep{Abbott16}, the scientific community has long been invested in studying the effects of cosmic structure on the observed signal \citep[see, e.g., one of the first examples ][]{Wang1996}. Here, we choose to focus on gravitational magnification, i.e. the enlargement of a source in the image plane of an observer due to the converging effect of one or more gravitational lenses along the line of sight.

For point-like electromagnetic (EM) sources this corresponds to an increase in brightness of a factor $\mu$ which has been shown to greatly affect the bright end of the luminosity functions of high redshift quasars and submillimeter galaxies \citep[e.g.,][]{Negrello2010,Wyithe2002}. Similarly, in the case of standard candles with known luminosity \citep[e.g., Type Ia supernovae or SNIa,][]{Nomoto1997}, magnification can induce a bias in the recovered distance-redshift relation. However, because an average null magnification is expected for each redshift bin, this bias is usually alleviated by flux-averaging multiple sources \citep{Wang2000}.

For gravitational wave mergers, previous works \citep[e.g.,][]{Dai2017, Oguri2018, Smith2018} have already studied the effects of lensing on a range of source population models and confirmed that, in the presence of a sharp cut-off in the intrinsic distribution, the observed one is smoothed out and transformed into a long and highly suppressed tail. More specifically, \citet{Broadhurst2018} claimed that a considerable fraction of LIGO-Virgo events to date might belong to this tail and that another sign of strong lensing, i.e. multiple images originating the same source, might have already been detected \citep{Broadhurst2019}. 

While this idea offers an explanation for the present-day tension with binary evolution models \citep[see, e.g., ][]{Dominik2012} that predict lower masses than observed, it is not favored by the data itself \citep{Hannuksela2019, 2019arXiv191003601S}. Furthermore, the tension it tries to explain might also be alleviated through tweaks to stellar evolution models \citep{Abbott2016b}.

The goal of this short paper is to offer some quantitative insights into the effects of lensing on the expected rates of gravitational wave mergers and highlight its general low likelihood.\footnote{In the interest of reproducibility, a Jupyter notebook offering a guided version of this work is available at {\url{https://www.github.com/contigiani/lensingGW}}.} This is done in light of the aforementioned claims and the proposed use of gravitational merger events as powerful standard sirens \citep{nissanke2013determining, Abbott2017}. 
In Sec.~\ref{sec:lensing} we discuss magnification effects on the measured GW signal and compare them to the EM case, while in Section~\ref{sec:rates} we derive the impact on the observed rates. In general relativity, light and gravitational waves move along the same geodesics. Because of this, the difference between the two can only be due to the dependence of the inferred source properties on $\mu$ and how \emph{efficiently} this dependence is translated into the observed rates. In Section~\ref{sec:results} we discuss our results and, finally, in Sec.~\ref{sec:conclusions} we draw our conclusions. 

\section{Lensing}
\label{sec:lensing}

\begin{figure}
\includegraphics[width=0.46\textwidth]{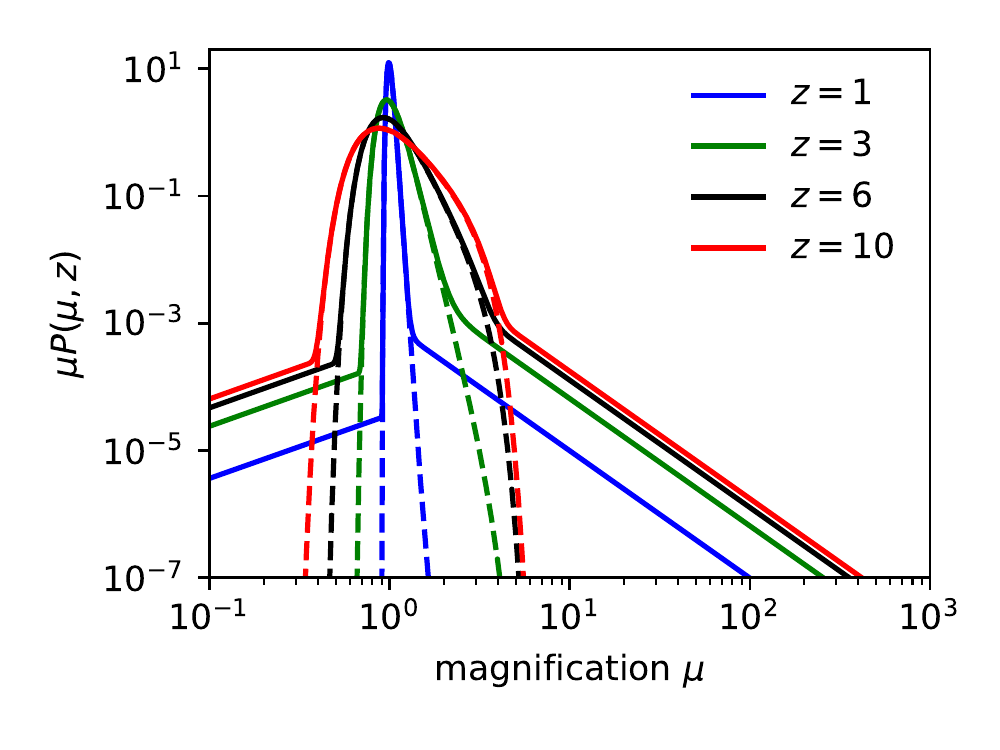}
\caption{PDF of the log-magnification ($\log \mu$) for different source redshifts $z$ used here. This figure is a rough approximation of more precise results, see e.g. figure 2 of \citet{Oguri2018}. While the weak lensing component of this distribution (dashed line) can be well approximated by assuming a log-normal distribution of the convergence, our power-law assumption for the strong lensing component underestimates this probability for $\mu\lesssim1$ and $\mu \sim 2$.}
\label{fig:Pmu}
\end{figure}

The value of the magnification $\mu$ for cosmological sources at various redshifts $z$ is modelled by a probability function $P(\mu, z)$ which can be obtained numerically by performing ray-tracing simulations \citep[e.g., ][]{Hilbert2007, Takahashi2011}. To simplify the notation, here we call $P(\mu)$ what is sometimes called $\frac{dP}{d\mu}$ in the literature. This quantity measures the distribution of magnification for all possible images of a given source and due to conservation of photons/gravitons on average we have null magnification,
\begin{equation}
    \langle \mu \rangle = \int d \mu \; \mu P(\mu, z) = 1.
    \label{eq:one}
\end{equation}    
More details about how this distribution should be interpreted are available in appendix A of \citet{Oguri2018}.

For this paper, we will use a simplified model of $P(\mu, z)$, calculated as the sum of two components:  weak and strong lensing. For the first, we assume a log-normal distribution for the convergence $\kappa$ \citep[as in, e.g.,][]{Taruya2002, Hada2018} and derive the corresponding magnification pdf using the relation:
\begin{equation}
    \mu \simeq \frac{1}{(1-\kappa)^2}.
\end{equation}
While this relation for $\mu$ and $\kappa$ is valid only in the limit of null shear $|\gamma| = 0$, it has been shown to accurately reproduce the weak lensing component of the magnification distribution \citep{Takahashi2011}, where $\kappa \lesssim 1$.
For the strong-lensing component, we do not assume any relation between $\mu, \kappa, |\gamma|$ and instead impose a power-law $P(\mu, z) \propto \mu^{-3}$ for $\mu>1$, calibrated empirically using the lensing depths of \citet{Oguri2018}. Finally, to simulate the demagnification tail, we assume a constant value for $\mu < 1$. The complete result is presented and discussed in Fig.~\ref{fig:Pmu}. In this work, we do not consider sources with $z>10$.

For EM sources, in the presence of magnification, the source flux is amplified by a factor $\mu$. If the redshift to the source is known and a cosmology is assumed, the result is a mismatch between the inferred luminosity ($\mathcal{L}$) and the intrinsic one ($\mathcal{L}_\ast$):
\begin{equation}
    \frac{\mathcal{L}}{\mathcal{L}_\ast} = \mu;
    \label{eq:L}
\end{equation}
while if only the luminosity is known (i.e. for standard candles) then the result is a mismatch between the inferred and true luminosity distance to the source:
\begin{equation}
    D(z) = \frac{D(z_\ast)}{\sqrt{\mu}},
    \label{eq:dist}
\end{equation}
where we call $z$ and $z_\ast$ the inferred redshift and the true one, respectively. We also refer to the corresponding luminosity distances as $D$ and $D_\ast$. The Jacobians of the transformations in Eqs.~(\ref{eq:L}) and (\ref{eq:dist}) are:
\begin{equation}
    \frac{\partial \mathcal{L}_\ast}{\partial \mathcal{L}} = \frac{1}{\mu}, 
\end{equation}
and 
\begin{equation}
    %\frac{\partial z^\prime}{\partial z} = \frac{\frac{\partial d_L}{\partial z} (z^\prime)}{\frac{\partial d_L}{\partial z}(z)} \sqrt{\mu}.
    \frac{\partial z_\ast}{\partial z} = \frac{D^\prime(z)}{D^\prime(z_\ast)} \sqrt{\mu}.
    \label{eq:SNJ}
\end{equation}

In the case of GWs, we limit ourselves to the inspiral phase of compact binary mergers. In this phase, the gravitational wave strain amplitude as a function of time, $h(t)$, carries information about both the distance of the source and the associated masses. The frequency evolution of the signal can be used to extract the redshifted chirp mass (an effective combination of the masses involved in the merger):
\begin{equation}
    \dot{f} \propto \mathcal{M} (1+z),
\end{equation} 
while the amplitude is connected to the inverse of the luminosity distance:
\begin{equation}
    h(t) \propto \frac{ A\left(\mathcal{M}(1+z)\right)}{D(z)},
\end{equation} 
where $A(-)$ is a function of the redshifted chirp mass alone. From here, it should be clear that both $\mathcal{M}$ and $D(z)$ can be extracted from the signal. 

In the presence of magnification, the observed strain is multiplied by a factor $\sqrt{\mu}$, and the mismatch between the intrinsic properties ($z_\ast, \mathcal{M}_\ast$) and the inferred ones ($z, \mathcal{M}$) is such that
\begin{equation}
    D(z) = \frac{D(z_\ast)}{\sqrt{\mu}},
    \label{eq:dl}
\end{equation} and
\begin{equation}
    \mathcal{M} = \mathcal{M}_\ast \frac{1+z_\ast}{1+z}.
    \label{eq:M}
\end{equation}
For $\mu>1$ this implies that distant events are assumed to be closer and more massive than they actually are, just like magnified electromagnetic sources are assumed brighter. An essential difference between the two, however, is that the dependence on magnification is significantly weaker for the GW merger parameter $\mathcal{M}$ compared to the luminosity $\mathcal{L}$,
\begin{equation}
    \frac{\mathcal{M}}{\mathcal{M}_\ast} = \frac{1+z_\ast}{1+z} \propto \mu^{s(z)},
    \label{eq:scaling}
\end{equation} with $s(z) < 0.5$ for any $z<z_\ast$ and $s(z) \to 0.5$ for increasing $z$. This can be easily shown by combining Eq.~(\ref{eq:dl}) and (\ref{eq:M}), together with the fact that the luminosity distance can be expressed, in a flat background, as the product of $(1+z)$ and a strictly increasing function of $z$ (comoving distance). 

To conclude this section, it is useful to point out that the Jacobian of the transformation $(\mathcal{M}, z) \leftrightarrow (\mathcal{M}_\ast, z_\ast)$ can be written as
\begin{equation}
    \frac{\partial \mathcal{M}_\ast}{\partial \mathcal{M}} \frac{\partial z_\ast}{\partial z} = \frac{D^\prime(z)}{D^\prime(z_\ast)} 
    \frac{1+z}{1+z_\ast}
    \sqrt{\mu}.
\end{equation}

\section{Rates}
\label{sec:rates}

\begin{figure*}
    \includegraphics[width=0.47\textwidth]{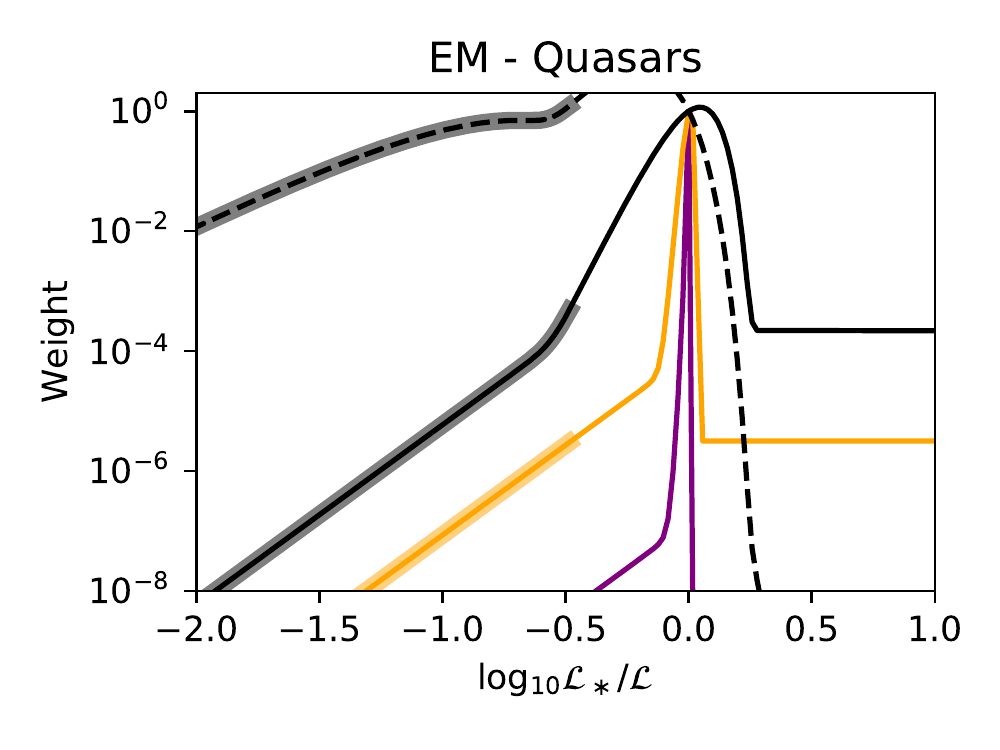}
    \includegraphics[width=0.47\textwidth]{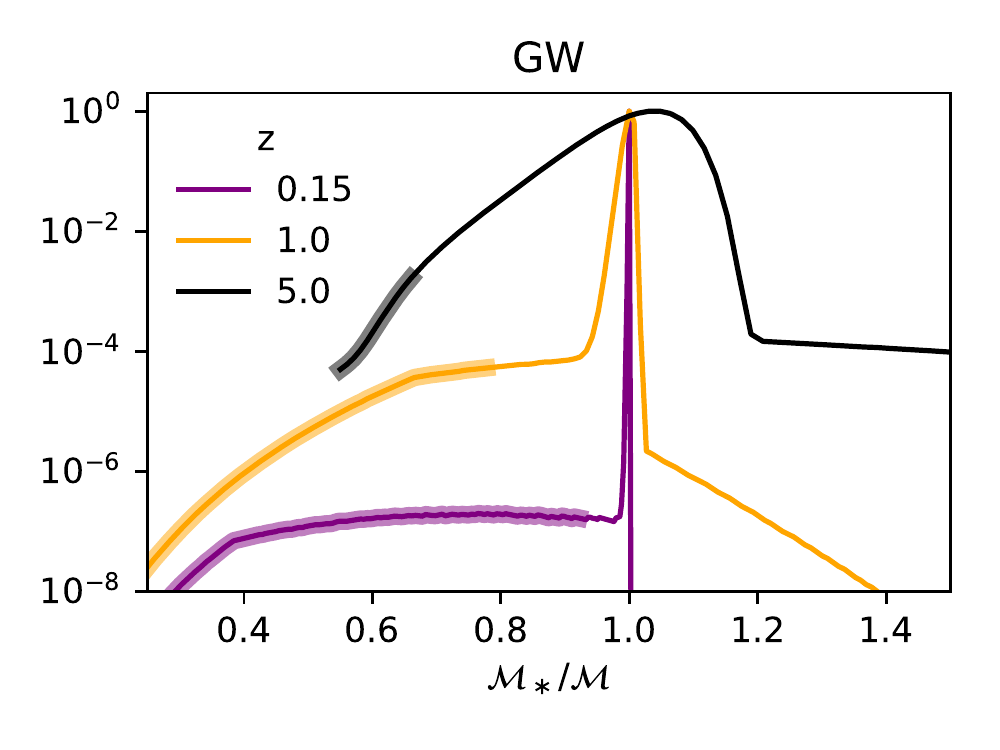}
    \caption{Relative contribution to the observed counts of transient GW events with chip-mass $\mathcal{M}$ or permanent EM sources with luminosity $\mathcal{L}$ by events with different intrinsic properties ($\mathcal{M}_\ast$ or $\mathcal{L}_\ast$). The observed redshift $z$ is equal to the intrinsic one for the EM case. The filled lines assume a  flat mass/luminosity function ($\mu>3$ for the thick lines), while the dashed line corresponds to the weights for an observed luminosity located well past the knee of a Schechter luminosity function ($\mathcal{L}/\mathcal{L}_{0} = 10$). For non-flat distributions, the relative contribution is the mass (or luminosity) function multiplied by these weights (see Eqs.~\ref{eq:GWE} and \ref{eq:QE}).}
    \label{fig:mass}
\end{figure*}

We write the observed rate of merger events per unit redshift and unit chirp mass as:
\begin{equation}
    r^{(GW)}(\mathcal{M}, z) = \frac{R(\mathcal{M}, z)}{1+z} \mathcal{E}_L^{(GW)}(\mathcal{M}, z),
\end{equation}
where $R$ is the intrinsic rate in the source frame, and $\mathcal{E}_L$ the lensing boost. Here, we separate the rate in two components:
\begin{equation}
    R(\mathcal{M}, z) = R(\mathcal{M})R(z),
\end{equation}
and, for the redshift-dependent part, we assume a rate which is proportional to the product of the comoving volume boosted by a factor $(1+z)^\beta$:
\begin{equation}
    R(z) \propto \frac{dV_c}{dz} (1+z)^\beta \propto \frac{d_L^2(z)}{E(z)} (1+z)^{\beta-2},
\end{equation}
where we use a standard $\Lambda$CDM cosmology with $E(z) = \sqrt{0.3 (1+z)^3 + 0.7}$. This power-law behaviour is expected if the merger rate of compact binary objects traces the star formation history \citep{Madau1998} at low redshift \citep{Dominik2013}. In this toy model, we also invert the sign of the power-law index $\beta=2.3$ at $z=2$, in order to simulate a peak in the star formation rate. 

Similar expressions can also be written for the rates of SNIa and the number counts of quasars:
\begin{equation}
    r^{(SN)}(z) = \frac{R(z)}{1+z} \mathcal{E}_L^{(SN)}(z),
\end{equation}
\begin{equation}
    n^{(Q)}(\mathcal{L}, z) = N(\mathcal{L}) \mathcal{E}_L^{(Q)} (\mathcal{L}, z).
\end{equation}    
Even though we assume that the intrinsic luminosity function of quasars $N(\mathcal{L})$ is not redshift dependent, lensing effects introduce this dependence in the observed $n(\mathcal{L}, z)$. The lensing boost factors can then be written as:
\begin{equation}
    \mathcal{E}_L^{(GW)} = \int d\left(\mathcal{M}_\ast/\mathcal{M} \right)\;
    \frac{R(\mathcal{M}_\ast)}{R(\mathcal{M})} W^{(GW)}(\mathcal{M}_\ast/\mathcal{M}, z),
    \label{eq:GWE}
\end{equation}
\begin{equation}
    \mathcal{E}_L^{(SN)} = \int d D_\ast \; 
    \frac{R(z_\ast)}{R(z)} W^{(SN)}(D_\ast/D, z),
\end{equation}
\begin{equation}
    \mathcal{E}_L^{(Q)} = \int d\log_{10} \left( \mathcal{L}_\ast/ \mathcal{L} \right)  \;
    \frac{N(\mathcal{L}_\ast)}{N(\mathcal{L})} W^{(Q)}(\mathcal{L}_\ast/\mathcal{L}, z),
    \label{eq:QE}
\end{equation}
where we have introduced the weight functions $W^{X}$, quantifying the contribution to the observed rates at $z, \mathcal{M}, \mathcal{L}$ from lensed events. These weight functions can be written as the product of the following terms. 
\begin{itemize}
    \item A lensing term. For each $z_\ast, \mathcal{M}_\ast$ and $\mathcal{L}_\ast$ there is an associated lensing probability. For GW and SN this is $P(\mu, z_\ast)$ because the measured redshift $z$, inferred from the luminosity distance, is different from the source redshift $z_\ast$. For the Q case this probability is simply $P(\mu, z)$ because it is measured directly. For $\mu>3$ we have $P(\mu, z_\ast)> P(\mu, z)$, meaning that we expect strong lensing to be particularly efficient for standard candles/sirens. Furthermore, because the expressions above are not written as integrals in $\mu$, this term also contains a probability volume, e.g. $d\mu /dz_\ast$ for the SN case. 
    \item A comoving volume term for the GW and SN cases. This is due to our assumption that $R(z)\propto dV_c$. Because lensing introduces contributions from a redshift range different from the observed $z$, a term $dV_c(z_\ast)/dV_c(z)$ is present. 
    \item A redshift evolution term for SN and GW. Similar to the previous case, except due to the assumed power-law dependence of $R(z)$. This term also accounts for the different redshifted rates and is equal to $\left(\frac{1+z_\ast}{1+z}\right)^{\beta-1}$. 
    \item A Jacobian term. As introduced in the previous section, the lensing transformation from intrinsic to observed quantities introduces an additional Jacobian factor.
\end{itemize}

 In the next section, we study in detail the impact of lensing magnification on the inferred chirp mass and redshift values and compare these results to the EM cases. We will work with the arguments of the integrals written above and, for ease of readability, we will also normalize these functions w.r.t. their value at null magnification ($\mu=1$). In particular, we chose not to focus extensively on the results of the integral $\mathcal{E}^{(GW)}_L$, since it strongly depends on the assumed mass function $R(\mathcal{M})$.  For accurate rates, we refer the reader to previous works \citep[e.g., ][]{Dai2017, Oguri2018, Broadhurst2018, Ng2018}.

\section{Results}
\label{sec:results}

    \subsection{Weight function}
    On the right side of Fig.~\ref{fig:mass}, we plot the contribution of different intrinsic chirp masses to the integral in Eq.~(\ref{eq:GWE}), while on the left-side we plot the equivalent result for light. These functions correspond to $W^{(GW)}$ and $W^{(Q)}$.
    
    The first obvious conclusion is that magnification affects more efficiently the inferred rates of GW mergers compared to EM sources at both high and low redshift. This is mainly because GW lensing gives access to a wider volume at higher redshift, corresponding to a higher Jacobian factor and significantly stronger lensing probabilities. These effects are the main discriminant between the two cases and are dominant at low redshift. 
    
    We note, however, that the GW weights are still low. If we focus on a LIGO-like source ($z\sim 0.15$), we see that, in order to have rates at mass $\mathcal{M}$ dominated by events at $\mathcal{M}_\ast\sim \mathcal{M}/3$, the mass function $R(\mathcal{M})$ should span roughly $7$ orders of magnitude between $\mathcal{M}_\ast$ and $\mathcal{M}$. While this has been shown to be possible, we point out that this roughly corresponds to a doubly-exponential tail, with 
    \begin{equation}
    R(M) \propto  e^{-e^{\mathcal{M}/\mathcal{M}_0}}
    \end{equation} and $\mathcal{M}_0 = \mathcal{M}_\ast$. This conclusion is mostly independent of our assumed mild redshift evolution.
    
    Despite the lower lensing weights for the EM case, we also show that a typical Schechter function $N(\mathcal{L}) \propto \exp(-\mathcal{L}/\mathcal{L}_\ast)/\mathcal{L}$ \citep{Schechter1976} is able to introduce a significant contribution from highly magnified sources at high $z$.
    
    \subsection{Lensing tail}
    \label{sec:tail}
    
    \begin{figure}
        \includegraphics[width=0.47\textwidth]{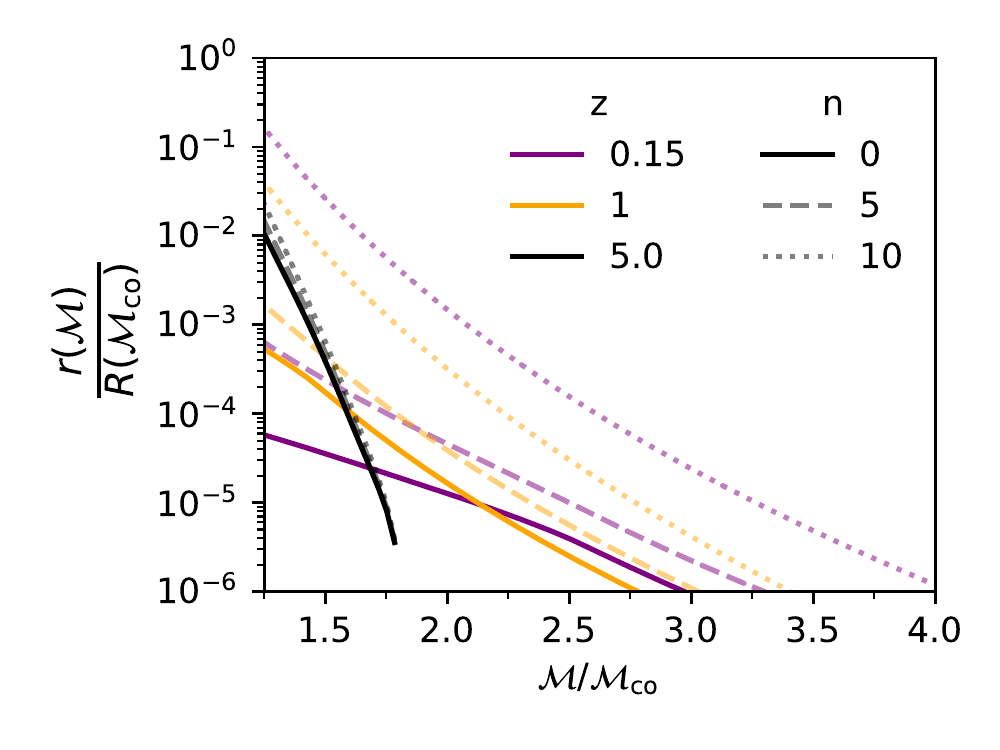}
        \caption{The shape of the lensing tail for truncated power-law distributions. The figure shows the observed rate $r(\mathcal{M})$ for an intrinsic chirp-mass function $R(\mathcal{M})\propto \mathcal{M}^{-n}$ truncated at $\mathcal{M}_{\mathrm{co}}$. The observed rate for $\mathcal{M}>\mathcal{M}_{\mathrm{co}}$ must therefore be due to lensed events. The dependence on $n$ is more striking for low inferred redshifts $z$ due to how the intrinsic chip masses are distributed in the volume at $z_\ast>z$. See Sec.~\ref{sec:tail} for more details.}
        \label{fig:truncation}
    \end{figure}    
    
    In Fig.~\ref{fig:truncation} we show the expected lensing tail of a truncated power-law distribution $R(\mathcal{M})\propto \mathcal{M}^{-n}$ for a few choices of $n$. Events measured with a chirp-mass larger than the cut-off value $\mathcal{M}_{\mathrm{co}}$ must be magnified mergers with intrinsic redshift $z_\ast>z$ and intrinsic chirp-mass $\mathcal{M}_\ast<\mathcal{M}$. 

    The prominence of this tail for a steep mass function (large $n$) and low redshift $z$ explains why a source distribution can be designed to produce a large number of lensed events \citep{Broadhurst2018}. It is useful to stress here that the main reason behind this is not the larger volume available to be lensed, but the fact that higher redshift events contributing to the low redshift rates are both more likely to be lensed and are also \emph{necessarily} located on a more abundant portion of the mass function. This is because the mapping $(\mathcal{M}, z) \leftrightarrow (\mathcal{M}_\ast, z_\ast)$ depends only on $\mu$. Despite the main advantage of amplifying the lensing tail compared to the naive expectation, this mechanism has the drawback of being efficient only for events with low $z$. For example, the shape of the $z=5$ lensing tail is less sensitive to the details of the mass function.
    
    Here we do not assume a lower limit for the values $\mathcal{M}_\ast$ and the integrals are truncated only because we impose $z_\ast<10$. While this choice is unrealistic, it is possible to verify that imposing a lower limit ${\mathcal{M}_\ast > 5~ \mathrm{M}_\odot}$ 1) does not affect the quantitative results of Fig.~\ref{fig:truncation} for $n<10$ and ${\mathcal{M}>20~ \mathrm{M}_\odot}$, and 2) has no impact on the qualitative results discussed in this section for all values of $n$.
    
    \subsection{Luminosity distance}
    Another consequence of the dependence of the observed mass $\mathcal{M}$ on the magnification $\mu$ is the broadness of the peak in Fig.~\ref{fig:mass}. The standard deviation of this distribution can be interpreted as an uncertainty in the measured $\mathcal{M}$, and, for an individual event it can be quite substantial: its value grows from $1$ to about $7$ per cent between $z=1$ and $z=5$. The main source of this scatter is the convergence distribution discussed in Sec.~\ref{sec:lensing} and it is not particularly affected by our chosen source redshift dependence $R(z)$. For a flat mass function, no significant bias is observed in this redshift range, meaning that the contributors to an event of observed chirp mass $\mathcal{M}$ and redshift $z$ are expected to have, on average, the same intrinsic properties.

\begin{figure}
        \includegraphics[width=0.47\textwidth]{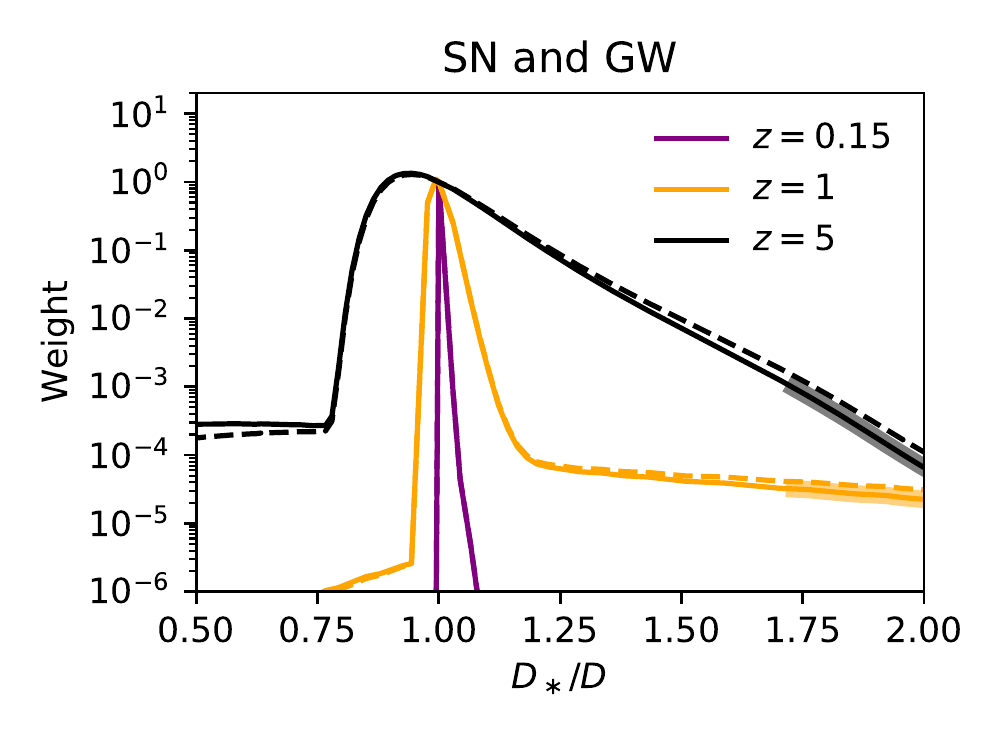}
        \caption{Relative contribution to the observed counts of transient GW events with redshift $z$ by events with a different intrinsic redshift $z_\ast$. This is plotted as a function of the direct observable, the luminosity distance $D(z)$. The filled lines correspond to a flat mass function ($\mu>3$ for the thick lines). The dashed lines represent the same result for the SN case, i.e. the case of transient EM sources of known luminosity for which lensing can also bias the result.}
        \label{fig:distance}
    \end{figure}

    In Fig.~\ref{fig:distance} we plot the equivalent of Fig.~\ref{fig:mass} for the luminosity distance $D(z)$. This is of particular interest because in the literature magnification effects are usually reported in terms of a smearing of the inferred distance instead of the inferred mass. For a flat mass function, we find a scatter of $2.5$ per cent at $z=0.15$ and $10$ per cent at redshift $z=5$, which is consistent with results from previous works \citep[e.g., ][]{Holz2005, Kocsis2006, Sathyaprakash2010, Oguri2016}. This value should be, however, compared to the present-day observational uncertainty in $D(z)$ of about $25$ per cent, dominated by the poorly constrained detector efficiency. 
    
    While not shown, one can also find that in the presence of a steep mass function, the inferred $D(z)$ is substantially more biased compared to the inferred $\mathcal{M}$. This is because $D_\ast$ and $\mathcal{M}_\ast$ scale differently with $\mu$ (Eq.~\ref{eq:dl} and \ref{eq:scaling}).

\section{Conclusions}
\label{sec:conclusions}
After studying the effects of gravitational lensing on the observed rates of GW mergers at low and high redshift, we conclude that magnification is not expected to significantly affect them. To show this, we have calculated the relative contribution of magnified events to the observed rates without assuming a specific chirp mass distribution. 

We have worked in the geometric optics limit to compare the effects of magnification on the observed chirp-mass function of GW mergers, and luminosity function for EM sources. Due to the larger wavelengths $\lambda$ of GWs, this approximation can break down if an object with a Schwarzschild radius comparable to $\lambda$ lies along the line of sight between source and observer \citep[see, e.g.,][]{2003ApJ...595.1039T}. The strength and rates of the resulting wave effects depend on the redshift and frequency considered \citep[see, e.g., for both ground-based and space-based detectors,][]{Sereno_2010, Dai_2018}. In all cases, however, these do not cause a direct bias in the parameters considered here due to the frequency-dependent signature they leave in the measured waveforms.

The LIGO-Virgo detector is currently on its third observing run, and in a few years it is expected to reach its design sensitivity. The expected statistical sample of mergers, made of hundreds or thousands of events, will allow a full reconstruction of the chirp mass distribution of the underlying populations. If the intrinsic distribution is extremely peaked, the observed one might be contaminated by highly lensed events with biased luminosity distances and chirp masses. However, not only this scenario is in conflict with the expectation from current stellar evolution models \citep[see, e.g., ][]{Belczynski16, Belczynski17}, but we have shown here that this would leave an easily recognizable signature in the LIGO rates due to 1) the wide range of probed masses at low redshift \citep{2016PhRvD..93k2004M} and 2) the flatness and low values of the lensing efficiency as a function of chirp mass (see Fig.~\ref{fig:mass}).

As an example, the contribution to mergers with an observed $\mathcal{M}\sim30$ $M_\odot$ and $z\sim0.15-1$ ($D \sim 700-1000$ Mpc) from events with lower $\mathcal{M}$ is suppressed by a factor $\sim10^6-10^4$. No matter how these lensed events are distributed in intrinsic chirp mass, the non-lensed events with similar properties should be both abundant and isolated from the highly suppressed lensing tail. These values roughly correspond to the $12$ mergers detected during the first and second observing run of LIGO-Virgo \citep{Abbott_2019}. In light of what is presented here, the absence of a larger number of events at $\mathcal{M} < 10$ M$_\odot$ (to which the detector has been shown to be sensitive), suggest that these events are not lensed.

These results offer guidance when interpreting magnification effects on the soon to be measured merger rates and are intentionally agnostic regarding detector or source population. The main conclusions hinge only on the weak dependence of the inferred binary properties on the factor $\mu$ and provide a general explanation for the established result that lensing contamination for luminosity-limited GW events are low for a wide range of detectors and source populations \citep[e.g.,][]{Sereno2010, Ding2015, Ng2018, Oguri2018}.

\section*{Acknowledgements}
The author acknowledges the helpful comments of the anonymous referee, and numerous discussions with Henk Hoekstra, Konrad Kuijken, and Samaya Nissanke. This research is supported by a de Sitter Fellowship of the Netherlands Organization for Scientific Research (NWO).

%%%%%%%%%%%%%%%%%%%%%%%%%%%%%%%%%%%%%%%%%%%%%%%%%%



\bibliographystyle{mnras}
\bibliography{bibliography}

% Don't change these lines
\bsp	% typesetting comment
\label{lastpage}
\end{document}